\documentstyle[12pt]{article}
\newcommand{\be}{\begin{equation}}
\newcommand{\ee}{\end{equation}}
\title{\bf
Generalized Coordinate Gauge, Nonabelian
 Stokes Theorem and the Dual QCD Lagrangian}
\author{ V.I.Shevchenko\thanks{shevchen@heron.itep.ru} and
Yu.A.Simonov\thanks{simonov@vxitep.itep.ru}.
\\
{\it Institute for Theoretical and Experimental Physics}\\
{\it 117218, B.Cheremushkinskaya, 25, Moscow, Russia}}
\date{}
\begin{document}
\maketitle
\vspace{1cm}
\centerline{\bf {Abstract}}
\vspace{3mm}
This is an extended version of the paper hep-th/9802134. 
Dual QCD Lagrangian is derived by making use of
the coordinate gauge of general type
where the 1-form (vector potential) is expressed as a contour integral of the
2-form (field strength) along an (arbitrary) contour $C$.
As another application a simple proof of the nonabelian Stokes theorem
is given.
\newpage

 The procedure of the gauge fixing is
an essential part of QCD \cite{ynd} and however final results do not
depend on the gauge, different forms of gauge conditions
are useful in different settings of physical problems.
For example, in high energy scattering in QCD the axial
gauge has proved to be useful \cite{lip}, while in the OPE analysis
\cite{sh} the Fock-Schwinger \cite{fss} (sometimes called the
coordinate or radial) gauge was applied (for
discussions and derivation see  \cite{sh1} and also \cite{dub}).

In another physical situation, where the time axis is
singled out, as e.g. in the heavy quarkonium theory,
the modified coordinate gauge \cite{bal} can be convenient.
This  gauge was used recently in the context of equations
for the quark \cite{sim1} and gluon \cite{sim2} Green's functions,
displaying the property of chiral symmetry breaking and confinement.

There is another set of studies where an emphasis is made on
formulation of gauge theory without gauge-dependent degrees
of freedom from the very beginning, and the role of dynamical
variables is played by 2-forms \cite{mand}, loop variables \cite{mm}
or some auxilary (vector - type) variables \cite{mit}.
These completely gauge invariant approaches encountered their own
difficulties and as a matter of fact many gauge-invariant observables
are easier to calculate using gauge dependent diagrammatic
rules.

Both in the coordinate gauge \cite{fss} and in its modified form
\cite{bal} the shape of the contour $C(x)$, in the integral,
connecting vector potential and the field strength,
\be
A_{\mu}(x) = \int\limits_{C(x)} dz_{\nu} {\alpha}_{\rho\mu}(z)
F_{\nu\rho}(z)
\label{stoks1}
\ee
is fixed and consists of straight lines. Inessential for
physical results, it may be inconvenient in the course
of computations. In particular, in the confining phase of QCD,
when the QCD string is formed between two colour charges
it would be advantageous to choose the contours $C$ lying on the
world sheet of the string; in this case one could do simplifying
approximations as in \cite{sim1,sim2}, namely to keep only
Gaussian field correlator. The decoupling of ghosts,
known to occur for the gauges (\ref{stoks1})
(see \cite{leo}, \cite{kor} and references therein) is also an attractive
feature, which suggests to look for generalizations of
(\ref{stoks1}) with arbitrary contours $C$.

There is also a deeper reason for the interest to the gauges
of the type (\ref{stoks1}).
It lies in the fact, that
(\ref{stoks1}) can imply dynamical connection between
field variables and geometrical (i.e. contour) variables.
We briefly outline this possibility at the end
of the present paper.

The gauge condition of the type we are
interested in was introduced for the first time in \cite{kor}.
In the present paper we give a refined treatment of this gauge,
paying special attention to some important
details, missing in the original paper.
Let us briefly mention them.
To define this gauge condition correctly,
the set of contours $C$, determining the gauge must satisfy some
additional requirement (eq.(\ref{auto}) of the present paper).
This condition is essential for the representation (\ref{stoks1})
to hold true.
With this requirement we are also able to
formulate
the gauge condition in the local form (eq.(\ref{gauge}) of
the present paper).
An immediate use of the generalized contour gauge
which also has not yet been discussed in the literature
is the ability
to give a short and direct proof of the nonabelian Stokes
theorem \cite{halpern,aref} as we do below in this paper.
Another line of development pursued below is the derivation
of the QCD Lagrangian in terms of the dual vector potential
and contour variables.

Let us proceed with the definition of the generalized contour gauge.
Let $M$ be a $d$-dimensional connected Euclidean manifold.
We choose some subspace $M_0$, $M_0 \subset M$
which
in general may be disconnected and of lower dimension than $d$.
In the simplest case, to be considered below,
$M_0$ consists of the only one point $x_0$.

For each point $x \in M\setminus  M_0$ we define the unique smooth contour
$C_{x_0}^{x}$, $x_0 \in M_{0}$ connecting points $x$ and $x_0$.
The contours are parametrized as follows:
\be
C_{x_0}^{x} :\;  z^{\mu} = z^{\mu}(s,x);
\> s\in[0,1];\> z^{\mu}(0,x) = {x_0}^{\mu};\>
z^{\mu}(1,x) = x^{\mu}
\ee
The map $M\setminus  M_0 \to M_0$ defined above is
naturally extended to $M \to M_0$ by setting
$C_{x_0}^{x_0}$ to be the unit contour: $z^{\mu}(s,x_0)\equiv {x_0}^{\mu}$.
The resulting map $M \to M_0$ is assumed to be smooth.
In the particular case when $M_0$ consists of the only point $x_0$
it means, that the manifold $M$ should be contractible.
This requirement plays an essential role in what follows.

Let us choose two arbitrary points
$z_{\mu}(s,x)$ and $z_{\mu}(s',x)$
on the given contour $C$
in such a way that the point $z_{\mu}(s',x)$
lies between points $z_{\mu}(s,x)$ and $z_{\mu}(1,x)=x_{\mu}$
(if $s$ is natural parameter, it simply
means that $s<s'$). We
assume the following condition - for any $s,s'$ there exists
$s''$ such that
\be
z_{\mu}(s,x) = z_{\mu}(s'',z(s',x))
\label{auto}
\ee
The geometrical meaning of (\ref{auto}) is simple:
for any point $z$ lying on some contour $C_{x_0}^{x}$
its own contour $C_{x_0}^{z}$ coincides with the corresponding part
of the contour $C_{x_0}^{x}$.
The eq.(\ref{auto}) does not mean, generally speaking, that
contours
$C_{x_0}^{x_1}$ and $C_{x_0}^{x_2}$
from different points $x_1 \neq x_2$
have no common points except $x_0$.
The condition that contours $C_{x_0}^{x}$ should not selfintersect
(the only condition discussed in \cite{kor}) is
necessary but not sufficient
to guarantee (\ref{auto}) and therefore to derive (\ref{rota})
and (\ref{gauge}) below.
The defined set of contours forms an oriented tree graph without
closed cycles according to (\ref{auto}).

Let us now start with the gauge potential $A_{\mu}(x)$ taken in some
arbitrary gauge and perform the gauge rotation
\be
A_{\mu}'(x) = {\Omega}^{+}(x) A_{\mu}(x) {\Omega}(x) + \frac{i}{g}
\> {\Omega}^{+}(x) {\partial}_{\mu} {\Omega}(x)
\label{ga}
\ee
where
$$
{\Omega}(x) = U(x,x_0) = Pexp{(ig\int\limits_{x_0}^{x} A_{\mu}(z) dz_{\mu})}
$$
and integration goes along the contour $C_{x_0}^x$.
The important point is the differentiation of the phase factors
\cite{mand,mm} which is a well defined procedure for our choice
of contours since the function $z_{\mu}(s,x)$ is given.
The contour derivative reads:
$$
{\partial}_{\mu} {\Omega}(x) = ig A_{\mu}(x) {\Omega}(x)
- ig \Omega(x) A_{\rho}(x_0)\>
\frac{\partial x_0^{\rho}(x)}{\partial x_{\mu}} +
$$
\be
+ ig \int\limits_0^1 ds \frac{\partial z_{\nu}(s,x)}{\partial s}
{\alpha}_{\rho\mu}(z) U(x,z(s)) F_{\rho\nu}(z(s))
U(z(s),x_0)
\label{dif}
\ee
where
\be
{\alpha}_{\rho\mu}(z) = \frac{\partial z_{\rho}(s,x)}{\partial x_{\mu}}
\ee
By $x_{0}(x)$ in the second term
at the r.h.s. of (\ref{dif})
we denote the initial point $x_0$ for the contour with the end point $x$.
This term is trivially absent if $M_{0}$
consists of the only one point.

Substituting (\ref{dif}) into the (\ref{ga}) one gets:
$$
A_{\mu}'(x) =
A_{\rho}(x_0)\>
\frac{\partial x_0^{\rho}(x)}{\partial x_{\mu}} +
$$
\be
+ U(x_0,x)
\int\limits_0^1 ds \frac{\partial z_{\nu}(s,x)}{\partial s}
{\alpha}_{\rho\mu}(z) U(x,z(s)) F_{\nu\rho}(z(s))
U(z(s),x_0)
\label{kro}
\ee
Taking into account the condition (\ref{auto})
and  the gauge transformation property
$
U(x_0,x)U(x,z(s)) F_{\rho\nu}(z(s))
U(z(s),x_0) \to {F}_{\rho\nu}'(z(s))
$
we arrive to the final result
\be
A_{\mu}'(x) =
A_{\rho}(x_0)\>
\frac{\partial x_0^{\rho}(x)}{\partial x_{\mu}} +
\int\limits_0^1 ds \frac{\partial z_{\nu}(s,x)}{\partial s}
\> \frac{\partial z_{\rho}(s,x)}{\partial x_{\mu}}
\> F_{\nu\rho}'(z(s))
\label{rota}
\ee
This formula was proposed in \cite{kor} and used without
derivation in \cite{sim2}.
In the rest of the paper we take $x_0$ to be a unique point
for all contours and
therefore $\partial x_0^{\rho} / \partial x_{\mu} = 0 $.

The eq.(\ref{rota}) leads to important local
condition for vector-potential.
To this end, note, that
solving (\ref{auto})
with respect to $s'$ we find:
$$
s' = f(s,s'',x);\;\; f(s,s,x) = 1
$$
Substituting $s'=f(s,s'',x)$ into (\ref{auto}) and differentiating
with respect to $s$ one gets:
\be
\frac{\partial z_{\mu}(s,x)}{\partial s} =
\frac{\partial z_{\mu}(s'',z(s',x))}{\partial
z_{\rho}(s',x)}\frac{\partial z_{\rho}(s',x)}{\partial s'}
\frac{\partial f(s,s'',x)}{\partial s}
\label{didi}
\ee
Then putting $s'$ to be equal to unity and
multiplying both sides of (\ref{rota}) by
$t_{\mu}(x) = (\partial z_{\mu}(s,x)/\partial s)_{s=1}$ we get
$$
A_{\mu}'(x)\cdot t_{\mu}(x) =
$$
$$
=
\int\limits_0^1 ds \frac{\partial z_{\nu}(s,x)}{\partial s}
\> \frac{\partial z_{\rho}(s,x)}{\partial x_{\mu}}
\left.\left(\frac{\partial z_{\mu}(s,x)}{\partial s}\right)\right|_{s=1}
\> F_{\nu\rho}'(z(s)) =
$$
\be
= \int\limits_0^1 ds \frac{\partial z_{\nu}(s,x)}{\partial s}
\> \frac{\partial z_{\rho}(s,x)}{\partial s}\> {(g(s,x))}^{-1}
\> F_{\nu\rho}'(z(s)) = 0
\label{ort}
\ee
where $g(s,x) = (\partial f(s,s'',x) / \partial s )_{s''= s}$.
The second equality holds by virtue of (\ref{didi})
and the third due to antisymmetry of $F_{\rho\nu}'$.

The condition (\ref{ort}) can be easily understood
taking into account, that phase factors along the contours $C_{x_0}^{x}$ specifying
the gauge are equal to unity:
$$
U(x,x_0) = Pexp{(ig\int\limits_{x_0}^{x} A_{\mu}'(z) dz_{\mu})} =1
$$
Since (\ref{ort}) holds for all $x$, one gets:
\be
A_{\mu}'(x) \> t_{\mu}(x) = 0
\label{gauge}
\ee

Specific examples of the gauge condition discussed in the present
article are known in the literature.
These are the radial or Fock-Schwinger gauge
\cite{fss} and its modified forms
\cite{bal,sim2}.
Note that due to the topological restrictions
stated above these gauge conditions
are always defined in some neighbourhood of their
origins
but might not be well
defined globally, in particular, in the case of topologically
nontrivial $M$.
This was noticed in different respect
also in {\cite{leo}}.

As an illustrative example let us consider the use of the
generalized gauge
condition for
the nonabelian Stokes theorem.
There are different proofs of this
theorem in the literature
\cite{halpern,aref}, but what
we are going to present is perhaps the simplest one.
It is close in spirit to the
paper \cite{halpern}. 
Namely, we define the gauge condition
in such a way that potential $A_{\mu}(x)$ on the contour is
expressed as a
function of field strength $F_{\mu\nu}(u)$ defined on the (arbitrary)
surface, bound by the contour. Then rewriting gauge-invariant Wilson
loop in this gauge we obtain a relation, valid in the chosen
gauge and as the last step put it into gauge-covariant form.
It was done in \cite{halpern} for the completely
fixed axial gauge condition, which is a convenient choice
in two dimensions (or for planar surfaces in higher
dimensional case). Our procedure allows one to
choose an arbitrary surface $S$ bound by the
simple contour $C = \partial S$ and therefore the gauge
condition we use entirely depends on the shape of $S$.
We parametrize the surface
 by $w^{\mu}(s,t);\> s,t \in [0,1] $.
and choose an arbitrary point ${x_0}^{\mu}$ on the surface
 in such a way, that
$w^{\mu}(0,t) \equiv {x_0}^{\mu}$.
If $s=1$ then  $w^{\mu}(1,t)$ goes along the contour $C$
and $w^{\mu}(1,0) = w^{\mu}(1,1)$ according to $\partial C = 0$.

The following important remark is in order.
It is usually assumed that $S$ has the disk topology, 
and the contour $C=\partial S$ is unknotted, in this
simplest case we are free in our choice of $M_0$, which may
consist of only one point, what we actually have used.
For this topology
it is always possible to define a set of contours obeying
(\ref{auto}) by, for example, continuous
deformation of the planar disk with the radial contours.
But in topologically nontrivial cases 
the proof should be modified.
(see in particular \cite{jap}).
In the rest of the present paper
we are concentrating on the disk topology.

According to (\ref{rota}) the gauge potential $A_{\mu}(z)$ is related to
$F_{\mu\nu}(z)$ in the following way:
\be
A_{\mu}(z(s,t))\>=\>
 \int\limits_0^1 ds' \frac{\partial z_{\nu}(s',x(t))}{\partial s'}
\> \frac{\partial z_{\rho}(s',x(t))}{\partial x_{\mu}(t)}
\> F_{\nu\rho}(z(s',t))
\label{qq}
\ee
Equation (\ref{qq}) is actually nothing else than the Stokes theorem in its
infinitesimal form. It is well known that the generalization to
finite contours is nontrivial in the nonabelian case,
in particular the integral $\int_S F_{\mu\nu} d\sigma_{\mu\nu} $ depends on the
surface even if the contour $C=\partial S$ is closed.
But this integral does not enter by itself in the
nonabelian Stokes theorem. Instead
the quantity which should be considered here is a $P$-ordered
exponent $Pexp(ig\int\limits_C A_{\mu} dx^{\mu})$.

Substitution of (\ref{qq}) into the definition of the
$P$-exponent leads to the expression:
$$
Pe^{ig\int\limits_{C} A_{\mu}(x) dx_{\mu} } =
$$
$$
=  1 +  \sum\limits_{n=1}^{\infty} {(ig)^n}
\int ..\int d\sigma_{\mu\nu}(w^{(1)}(s_1,t_1)).. d\sigma_{\rho\phi}
(w^{(n)}(s_n,t_n))\>
$$
\be
 F_{\rho\phi}(w^{(n)}(s_n,t_n)).. F_{\mu\nu}(w^{(1)}(s_1,t_1))
\>\theta \left(t_1 > t_2 >..> t_n \right)
\label{sst}
\ee
Note the ordering procedure in (\ref{sst})
-- only the points along the contour $C$ are
ordered with respect one to another, i.e.
ordered in parameters $t_i$, while the
integrals over $s_i$ are taken independently for each $t_i$.

To bring (\ref{sst}) to the gauge covariant form
 we introduce phase factors along the $s$-direction on the surface,
 which are equal to unity due to (\ref{gauge}),
 i.e. we replace
$
F_{\mu\nu}(w(s,t)) \to G_{\mu\nu}(w(s,t)) =
U(x_0; w(s,t))F_{\mu\nu} (w(s,t)) U(w(s,t); x_0)
$
If the point $x_0$ does not lie on the contour $C$
the gauge-covariant answer reads:
\be
 Pe^{ig\int\limits_{{C}_{x^{*}x^{*}}} A_{\mu} dz_{\mu} } =
U(x^{*}, x_0)
{\cal P} e^{ig \int\limits_{S} d\sigma_{\mu\nu}(z) G_{\mu\nu}(z)}
U(x_0,x^{*})
\label{bre}
\ee
where the meaning of the ordering simbol $\cal P$ is explained
in (\ref{sst}).
Under the gauge rotations both sides of (\ref{bre}) are transformed
in the same way.
The more often used gauge-invariant form of (\ref{bre}) is
\be
Tr\> Pe^{ig\int\limits_{C} A_{\mu} dz_{\mu} } =
Tr\> {\cal P} e^{ig \int\limits_{S}
d\sigma_{\mu\nu}(z) G_{\mu\nu}(z)}
\label{qsi}
\ee
We stress again, that the exact meaning of the symbol ${\cal P}$
is completely determined by the choice of the set of contours,
defining the gauge which may be done as the most convenient one
for a given application of the nonabelian Stokes theorem.

It is worth noting that all formulas used above hold true in abelian
case too, the ordering operations are not necessary in this case.
In particular, for the given abelian gauge field strength $F_{\mu\nu}(x)$
one can define a set of contours $\{C(x,x_0)\}$ obeying all necessary conditions
discussed above and obtain a set of the gauge potentials
$\{A_{\mu}(x)\}$ by using formula (\ref{rota}).
Then all field configurations $\{ A_{\mu}(x) \}$ are gauge equivalent
as it is clear from (\ref{ga}). In other words one can say, that
variation of the contour's shape in (\ref{rota})
with the fixed $F_{\mu\nu}$ in abelian theory
leads only to gauge variation of the field $A_{\mu}(x)$ and therefore
the contour degree of freedom is not dynamical in this case,
all dynamics is encoded in the field strength tensor.
The situation changes however if we introduce monopoles into
the theory. The relation between field strength tensor and gauge
potential takes the form:
\be
F_{\mu\nu}(x) = \partial_{\mu} A_{\nu}(x)
- \partial_{\nu} A_{\mu}(x) + 2\pi g \sum {\tilde G}_{\mu\nu}(x)
\ee
where
$$
{\tilde G}_{\mu\nu}(x) = {\epsilon}_{\mu\nu\alpha\beta}
\int\limits_{\Sigma} d {\sigma}_{\alpha\beta}(y) {\delta}^{(4)}(x-y)
$$
and $\Sigma$ is an (arbitrary) surface, which is a worldsheet
of the Dirac string.
Then the contour $C(x)$ defining the gauge rotation $\Omega(x)$ in
(\ref{ga}) can be smoothly deformed until it intersects
$\Sigma$. If such intersection happens, gauge potential $A_{\mu}(x)$
receives a singular contribution proportional to
$$
\int\limits_{x_0}^{x} dz_{\nu}
\int\limits_{\Sigma} d {\sigma}_{\alpha\beta}(y)\>
\frac{\partial z_{\rho}}{\partial x_{\mu}}\>
{\epsilon}_{\nu\rho\alpha\beta}
{\delta}^{(4)}(z-y)
$$
It is seen, that even though Dirac strings are gauge degrees of freedom, the
contours $C$ may become dynamical ones in a theory with monopoles.
We leave this set of questions for the future publication and now
focus our attention on the dual form of the QCD
partition function.

There are different ways to integrate out gauge potentials
in nonabelian theories (see
for example \cite{mit}, \cite{halpern}, \cite{dual} and references therein).
The gauge condition discussed in the present
paper looks promising for that purpose
because it gives an explicit expression for
$A_{\mu}$ in terms of $F_{\mu\nu}$.
It should be noticed, that (\ref{ga})
indeed singles out a unique representative for any gauge orbit. To see this,
let us consider two gauge-equivalent connections $A$ and
$\tilde A = {\omega}^{+} A \omega + i/g\> {\omega}^{+}
{\partial} {\omega}$ and perform rotations (\ref{ga}):
$$
A' = {\Omega}^{+}[A] A\> {\Omega}[A] + \frac{i}{g}
{\Omega}^{+}[A] {\partial} {\Omega[A]}
$$
$$
{\tilde A}' = {\Omega}^{+}[\tilde A]\> \tilde A
{\Omega}[\tilde A] + \frac{i}{g}
{\Omega}^{+}[\tilde A] {\partial} {\Omega[\tilde A]}
$$
where integration in $\Omega[A]$ and $\Omega[\tilde A]$ goes along
one and the same set of contours. One easily finds, that
$ {\tilde A}'(x) = {\omega}^{+}(x_0) A'(x) {\omega}(x_0)$ where $x_0$
is the base point, i.e. resulting potentials coincide up to the global
gauge rotation.

Let us proceed with the complete gauge fixing.
Our strategy is a straightforward extension
of \cite{halpern}. We choose
the 4-bein $p_{\mu}^{i}(x)$; $\>\mu , i =1..4$ smoothly
depending on $x$:
$$
p_{\mu}^{i}(x) p_{\nu}^{i}(x) = {\delta}_{\mu\nu} \>; \;\;
p_{\mu}^{i}(x) p_{\mu}^{j}(x) = {\delta}^{ij}
$$
A set of orthogonal curves $x_{\mu}^{i} = x_{\mu}(s^i), \> i=1..4$
is defined in such a way that
$$
p_{\mu}^{i}(x) = \frac{d x_{\mu}(s^i)}{ds^i}
$$
The potential $A_{\mu}(x)={\hat A}^i(x) \>p_{\mu}^i(x)$
is taken to vanish on one of the chosen lines;
assuming this line to be defined by the conditions $x^i = const, \> i=1..3$
one gets:
$$
A_{\mu}(x(s^4)) = A_{\mu}(x^1,x^2,x^3)=0
$$
The piecewise contour $z_{\mu}(s,x)$ is then defined as follows:
\be
z_{\mu}(s,x) = x_{\mu}^{0} +
\sum\limits_{i=1}^{3} \int\limits_{s_0^i}^{s^i}
d{\hat s}^i p_{\mu}^i (x({\hat s}^i))
\label{qqq}
\ee
The path goes along the
lines $x^i$ consequtively with respect
to $i$, i.e. initial points in any integral from (\ref{qqq})
 coincide with
the end points for the previous integral: $x(s^{i-1})=x(s_0^{i})$.
It leads to the following local conditions for the potential:
\be
{\hat A}^3 =
 {\hat A}^1(x^2) = {\hat A}^2(x^1, x^3) = {\hat A}^4(x^1,x^2,x^3) =0
\label{ooo1}
\ee
The eq.(\ref{rota}) with the contour's definition
(\ref{qqq}) allow to express potential in terms of the field
strengths.

In the simplest particular case when $p_{\mu}^i = {\delta}_{\mu}^i$,
the expression (\ref{ooo1}) defines the so called completely
fixed axial gauge, used in \cite{halpern}.
One can easily check that the following conditions on the
usual coordinate components of the vector potential
hold true:
\be
A_3 = A_1(z_0) = A_2(x_0, z_0) = A_4(x_0, y_0, z_0) =0
\label{ooo2}
\ee
It is seen that it is a particular case
of generalized contour gauges discussed in this paper.

For completely fixed gauges
the following relation is very useful \cite{halpern}:
\be
\int D A \delta (CGF) \delta (G - F[A]) = \delta(I[G])
\label{bian}
\ee
where $I[G]$ is the Bianchi form
$I[G] = D_{\mu} {\tilde G}_{\mu\nu}$
and $\tilde G$ is the dual tensor ${\tilde G}_{\mu\nu}
= 1/2\> {\epsilon}_{\mu\nu\alpha\beta} G_{\alpha\beta}$.
Delta function $\delta(CGF)$ fixes gauge completely.
The expression (\ref{bian}) is local and Lorentz--covariant,
taking into account that locally (\ref{ooo1}) is equivalent to
(\ref{ooo2}) we conclude, that (\ref{bian}) works for
the generalized completely fixed gauges as well as for axial gauge.
It is straightforward then to integrate
 out vector potentials $A_{\mu}$ in the partition function.
The QCD partition function reads:
$$
Z[J] =\int DA D\psi D\bar\psi \delta(CGF)
$$
\be
\int DG \delta(G-F[A])
e^{ \int d^4 x \left(
-\frac{1}{4g^2}\>G_{\mu\nu}^2 + J_{\mu\nu} G_{\mu\nu}
+ \bar\psi (i{\gamma}_{\mu} D_{\mu}[A] + m) \psi \right)}
\ee
where the field strength tensor $F_{\mu\nu}[A] = {\partial}_{\mu} A_{\nu}
- {\partial}_{\nu} A_{\mu} -i [A_{\mu}A_{\nu}]$,
covariante derivative $D_{\mu}[A] = {\partial}_{\mu} - i A_{\mu}$
and $J_{\mu\nu}$ - external tensor source.

It is convenient to introduce a nonlocal operator
\be
K_{\mu}^{\nu\rho}(x,y) =
\int\limits_0^1 ds \frac{\partial z_{\nu}(s,x)}{\partial s}
\> \frac{\partial z_{\rho}(s,x)}{\partial x_{\mu}}
{\delta}^{(4)}(z-y)
\label{kkk}
\ee
Then the gauge fixing $\delta(CGF)$ implies due to (\ref{rota}) that
\be
A_{\mu}[G](x) =
\int d^4 y K_{\mu}^{\nu\rho}(x,y) G_{\nu\rho}(y)
\ee

By using (\ref{bian}) the partition function reads:
$$
Z[J] = \int D\psi D\bar\psi Dk_{\mu}
\int DG \; e^{i \int d^4 x \> k_{\nu}^a ( \partial_{\mu} {\tilde G}_{\mu\nu}^a
+ f^{abc} A_{\mu}^b[G]{\tilde G}_{\mu\nu}^c)}
$$
\be
e^{\int d^4 x \left(
-\frac{1}{4g^2}\>G_{\mu\nu}^2
+ J_{\mu\nu} G_{\mu\nu}
+  \bar\psi (i{\gamma}_{\mu} {\partial}_{\mu} + m) \psi
+ \bar \psi {\gamma}_{\mu} \psi A_{\mu}[G] \right) }
\label{zzz}
\ee
After the Gaussian $G$--integration one has:
$$
Z[J]= \int Dk_{\mu} D\psi D\bar \psi \left( det \Delta \right)^{-\frac12}
e^{ \int d^4 x
 (\bar\psi (i{\gamma}_{\mu} {\partial}_{\mu} + m) \psi)}
$$
\be
e^{- \int d^4 x \int d^4 y
 ( P_{\alpha\beta}^a (x)
+ J_{\alpha\beta}^a(x))
{\left(\frac{1}{\Delta}\right)}^{ab}_{\alpha\beta\rho\sigma}(x,y)
 ( P_{\rho\sigma}^b (y) + J_{\rho\sigma}^b(y))}
\label{zzz1}
\ee
where
\be
P_{\alpha\beta}^a (x) =
{\epsilon}_{\alpha\beta\mu\nu} \partial_{\mu} k_{\nu}^a(x)
+ \int d^4 w
\bar \psi (w) i {\gamma}_{\mu} t^a \psi (w)
 K_{\mu}^{\alpha\beta}(w,x)
\ee
and the propagator
\be
{\Delta}^{ab}_{\alpha\beta\rho\sigma}(x,y) =
\frac{1}{4g^2} {\delta}_{ab} {\eta}_{\alpha\rho}{\eta}_{\beta\sigma}
\delta(x-y) + i  f^{abc} {\epsilon}_{\xi\gamma\alpha\beta}
k_{\gamma}^c(x) K_{\xi}^{\rho\sigma}(x,y)
\label{delta}
\ee
Using abelian language
the second term in the r.h.s. of (\ref{delta}) represents the
"monopole current", while $k_{\mu}$ plays the role of the dual
vector-potential.

In the absence of quarks and external currents the "dualized"
partition function reads:
\be
Z = \int Dk_{\mu} e^{-S_{eff}[k] }
\ee
The effective dual action $S_{eff}[k]$ is similar to the one
obtained in \cite{halpern}:
\be
S_{eff}[k] = \frac12 Tr\>ln\> \Delta + \int d^4 x \int d^4 y\>
  {\hat F}_{\alpha\beta}^a[k] (x)
{\left({\Delta}^{-1} \right)}^{ab}_{\alpha\beta\rho\sigma}
  {\hat F}_{\rho\sigma}^b[k] (y)
\label{zzz3}
\ee
and notice that dual field strength is a linear function of dual potential
$k$ in our formulation:
$$
F_{\alpha\beta}[k] = \frac12 {\epsilon}_{\alpha\beta\mu\nu} (
{\partial}_{\mu} k_{\nu} - {\partial}_{\nu} k_{\mu})
$$

The stationary point ${\bar k}_{\mu}(x)$ of the effective action
can be found from the solution of the equation
\be
\left.\frac{\delta S_{eff} [k]}{\delta k(z)} \right|_{k=\bar k} = 0
\ee
In the large $N_c$ limit this yields in particular the effective
action for quarks of the following form:
\be
S_q = \int d^4 u\int d^4 w
\left( \bar \psi (u) i {\gamma}_{\mu} t^a \psi (u)
\left[{\left(  K{\Delta}^{-1} K\right) }_{\mu\nu}^{ab} (u,w)
\right]
\bar \psi (w) i {\gamma}_{\nu} t^b \psi (w)
\right)
\ee

and the kernel $K{\Delta}^{-1} K$
is to be evaluated at the stationary point ${\bar k}_{\mu}$.

The dual action is essentially nonlocal, which is
encoded in $\Delta(x,y)$. This property is common
to dualized nonabelian theories (see \cite{dual}).
But in our case we have an additional freedom --
to choose 4-bein $p_{\mu}^i$ in such a way, that the
operator $K(x,y)$ takes the most convenient form
for a given dynamical problem.
This way of research which is ideologically close to the
abelian projections method will be discussed elsewhere.

\bigskip

{\large\bf Acknowledgments}

The work was supported by the RFFI grants 96-02-19184a, 96-15-96740,
and partially by RFFI-DFG grant 96-02-00088G.
The authors are grateful to A.Levin, M.Olshanetsky, V.Novikov
for valuable discussions and to G.Korchemsky for pointing authors
attention to the Ref.\cite{kor}.

\end{document}